\documentclass[twocolumn]{aastex63}
\journalinfo{}
\accepted{2026 October 14}
\usepackage{amsmath}
\usepackage{amssymb}
\usepackage{natbib}
\usepackage{booktabs}

\newcommand{\msun}{M_\odot}
\newcommand{\alphatrue}{\alpha_\mathrm{true}}

\newcommand{\alphanaive}{\hat{\alpha}_\mathrm{naive}}

\newcommand{\fb}{f_\mathrm{b}}
\newcommand{\ftwin}{f_\mathrm{twin}}
\newcommand{\mobs}{m_\mathrm{obs}}

\newcommand{\mpri}{m_1}
\newcommand{\qmin}{q_\mathrm{min}}
\newcommand{\Ncross}{N_\mathrm{cross}}
\newcommand{\progenax}{\texttt{progenax}}
\newcommand{\fluxax}{\texttt{fluxax}}

\newcommand{\numpyro}{\texttt{NumPyro}}

\begin{document}

\title{Confidently Wrong: Why Ignoring Binaries Biases IMF Inference at Large Sample Sizes}

\author[0000-0003-4423-0660]{Anna L.\ Rosen}
\affiliation{Department of Astronomy, San Diego State University, 5500 Campanile Dr, San Diego, CA 92182, USA}
\affiliation{Computational Science Research Center, San Diego State University, 5500 Campanile Dr, San Diego, CA 92182, USA}
\email{alrosen@sdsu.edu}

\begin{abstract}
The stellar initial mass function (IMF) high-mass slope $\alpha$ is routinely measured by fitting single-star models to photometric samples that contain unresolved binaries. This practice introduces a systematic negative bias on $\alpha$ that is \emph{constant} with sample size $N$. Posterior credible intervals shrink as $1/\sqrt{N}$, so at sufficiently large $N$ the bias exceeds the full 95\% credible-interval width and the true value falls outside the credible interval --- a regime we call ``confidently wrong.'' Building on earlier studies of the bias magnitude, we quantify the sample size at which this bias exceeds the statistical uncertainty.
We quantify this bias using two idealized observation operators: \emph{mass-addition} ($\mobs = m_1 + m_2$), a formal upper bound on unresolved-system mass overestimation, and \emph{luminosity-addition} ($\mobs = L^{-1}(L_1 + L_2)$), an idealized lower-bias photometric case based on the ZAMS mass--luminosity relation. Across four astrophysical environments spanning $\alpha = 1.60$--$2.30$, we find:
(1) mass-addition bias of 0.054--0.086 with crossover to confidently wrong at $\Ncross \sim 5{,}000$--$10{,}000$;
(2) luminosity-addition bias of 0.011--0.021 with $\Ncross \sim 75{,}000$--$150{,}000$;
and (3) a binary-aware mixture likelihood that marginalizes over the \citet{moe2017} binary population model recovers the true slope in the synthetic tests presented here. Published single-star IMF slopes can therefore plausibly carry systematic errors of order 0.01--0.09 if unresolved binaries are not modeled, comparable to or exceeding reported uncertainties in some regimes. Since current and upcoming surveys (Gaia, JWST, Roman, LSST) will deliver large resolved-star samples, binary-aware inference is likely necessary to avoid binary-driven systematic bias in the large-$N$ single-star-fitting regime.
\end{abstract}


\section{Introduction} \label{sec:intro}

The stellar initial mass function (IMF), the distribution of stellar masses at birth, sets the supernova rate, chemical enrichment, and energy budget of star-forming galaxies \citep{kroupa2002, bastian2010}. Its high-mass slope $\alpha$, defined so that $\xi(m) \equiv dN/dm \propto m^{-\alpha}$ for $m \gtrsim 1\,\msun$ (Salpeter = 2.35 in this convention), is the single most consequential IMF parameter: it controls the ratio of massive to low-mass stars. Whether $\alpha$ is universal or varies with environment remains one of the oldest open questions in star and galaxy formation \citep{salpeter1955, kroupa2001, chabrier2003, bastian2010, hopkins2018, hennebelle2024}.

Unresolved binary stars contaminate every photometric IMF measurement. A substantial fraction of stars, from 22\% of M~dwarfs to 90\% of O~stars, reside in binary or higher-order multiple systems \citep{duchene2013, sana2012, moe2017, eldridge2022, sana2025}. When a binary system is unresolved, the observer infers a single ``system mass'' that is systematically higher than the primary mass. This shifts the observed mass function toward higher masses, mimicking a shallower (less negative) IMF slope. The problem extends beyond the IMF: unresolved multiples inflate inferred cluster masses \citep{weidner2009}, create apparently ultramassive stars above the true stellar upper-mass limit \citep{maizapellaniz2008}, and bias stellar parameters and abundances in spectroscopic surveys \citep{elbadry2018}.

\citet{kroupa1991} showed that unresolved binaries significantly bias the low-mass luminosity function, and \citet{maizapellaniz2008} showed that unresolved multiple systems and chance superpositions bias massive-star IMF determinations. \citet{weidner2009} found that even 100\% binarity shifts the observed high-mass slope by $\lesssim 0.1$. While tools such as BASE-9 \citep{stenning2016} can model unresolved binaries, ASteCA adopts a fixed binary fraction when generating synthetic clusters \citep{perren2015}. Building on these studies, we quantify the \emph{crossover sample size} $\Ncross$ at which the bias exceeds the credible interval width, as a function of observation operator (mass- or luminosity-based) and birth environment.

The \citet{moe2017} model provides primary-mass--dependent binary fractions and joint distributions of orbital period and mass ratio calibrated from solar-type to O stars and over nearly six orders of magnitude in orbital period. This framework makes it possible to compute the crossover sample size as a function of environment using empirically motivated binary statistics. The mass-dependent structure of \citet{moe2017} is essential because the contamination from unresolved binaries itself depends on the underlying IMF slope.

The critical question is not whether the bias exists --- prior work establishes that it does --- but at what sample size it begins to matter statistically. The systematic bias on $\alpha$ is approximately constant while statistical uncertainty shrinks as $1/\sqrt{N}$. We define the crossover sample size $\Ncross$ by comparing the absolute posterior-median offset with the full 95\% credible-interval width.

Current and upcoming surveys are already delivering sample sizes where this effect becomes unavoidable. Gaia DR3 contains $>10^5$ stars in nearby open clusters \citep{gaia2023}, and Gaia DR4 will provide ${\sim}10^6$ astrometric orbital solutions that directly constrain binary populations. The Vera C.\ Rubin Observatory's LSST, now commissioning, will deliver deep photometry of thousands of resolved stellar populations with $10^3$--$10^6$ stars each \citep{ivezic2019, usher2023}. Early Rubin commissioning observations already reveal unresolved binary sequences in 47~Tucanae \citep{cordoni2025rubin}. JWST is resolving individual stars in the outskirts of the Small Magellanic Cloud where binary contamination must be modeled to recover the intrinsic IMF \citep{legnardi2025}. The Nancy Grace Roman Space Telescope, scheduled to launch by ${\sim}2027$, will extend resolved-star censuses to dust-obscured young massive clusters, including the Arches, Quintuplet, and Galactic Center young stellar populations, where the IMF slope measurements remain actively debated and sample sizes will reach $N>10^4$. Because statistical uncertainties shrink as $1/\sqrt{N}$ while binary-driven bias remains approximately constant, these large surveys will inevitably enter the bias-dominated regime if binaries are ignored. In our benchmark tests this crossover occurs at $\Ncross \sim 5{,}000$--$150{,}000$, depending on the observation operator.

In this paper, we quantify the binary contamination bias using two idealized observation operators: \emph{mass-addition} ($\mobs = m_1 + m_2$) and \emph{luminosity-addition} ($\mobs = L^{-1}(L_1 + L_2)$). Both operators produce a bias-dominated regime at survey-relevant sample sizes, and a binary-aware mixture likelihood removes the bias. Section~\ref{sec:method} describes the binary population model and inference framework; Section~\ref{sec:results} presents the bias, crossover, and recovery results; and Section~\ref{sec:discussion} discusses implications for current and future surveys.

\section{Methods} \label{sec:method}

In this paper, we construct a forward model for unresolved binary contamination of the IMF high-mass slope and test parameter recovery across four astrophysical environments using Bayesian inference. We describe the IMF parameterization (\S\ref{sec:imf}), the binary population model (\S\ref{sec:binaries}), the two limiting observation operators (\S\ref{sec:obs}), and the inference framework (\S\ref{sec:inference}).

\subsection{IMF Model} \label{sec:imf}

We adopt the \citet{maschberger2013} IMF parameterization, a smooth three-parameter distribution interpolating between the low-mass turnover and the high-mass power law:
\begin{equation} \label{eq:maschberger}
    \xi(m) \propto \left(\frac{m}{\mu}\right)^{-\alpha} \left(1 + \left(\frac{m}{\mu}\right)^{1-\alpha}\right)^{-\beta}
\end{equation}
where $\alpha$ is the high-mass slope (the parameter of interest), $\mu$ is the characteristic mass ($\mu = 0.2\,\msun$), and $\beta$ governs the low-mass turnover ($\beta = 1.4$). This parameterization has two practical advantages over the piecewise power-law forms of \citet{kroupa2001}: (1) continuous derivatives enable efficient gradient-based sampling (HMC/NUTS) without divergent transitions near break masses, and (2) the normalization constant has a closed-form expression. For $m \gtrsim 0.5\,\msun$, where binary contamination is most significant, both parameterizations share the same high-mass slope $\alpha$, so the inferred bias on $\alpha$ is expected to be similar.

We fix $\mu$ and $\beta$ and treat $\alpha$ as the sole free parameter. This isolates binary contamination of the high-mass slope before introducing parameter degeneracies. Joint inference of all three Maschberger parameters is beyond the scope of this work (\S\ref{sec:limitations}).

We consider four values of $\alpha$ spanning the literature range ($1.60$--$2.30$), drawn from the \citet{marks2012} environment-dependent parameterization \citep[see also][]{jerabkova2018, hennebelle2024} and summarized in Table~\ref{tab:environments}. These labels denote benchmark IMF regimes; our results depend on $\alpha$, not on the physical mechanism that sets it. The metallicities are derived as $Z = Z_\odot 10^{[\text{Fe/H}]}$ with $Z_\odot = 0.02$ \citep{tout1996} and enter only through the ZAMS mass--luminosity relation for the L-add operator.

\begin{table}[h]
\centering
\caption{Benchmark environments used throughout this paper.}
\label{tab:environments}
\begin{tabular}{lcc}
\hline\hline
Environment & $\alphatrue$ & $Z$ \\
\hline
Solar neighborhood & 2.300 & 0.020 \\
Young massive cluster & 2.070 & 0.0063 \\
Low-$Z$ globular & 1.765 & 0.0006 \\
Low-$Z$ starburst & 1.604 & 0.00014 \\
\hline
\end{tabular}
\end{table}

\begin{figure}[t!]
\centering
\includegraphics[width=\columnwidth]{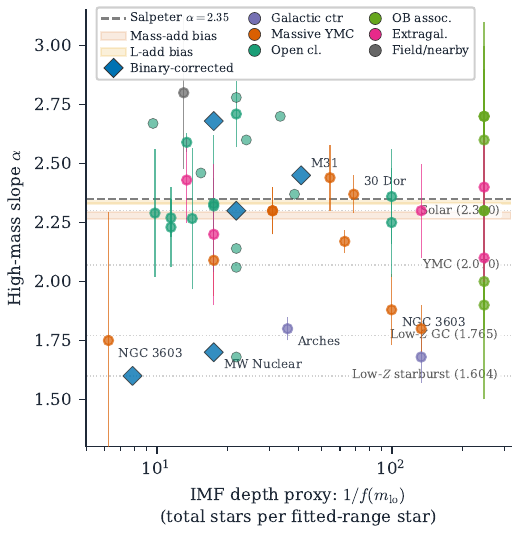}
\caption{Literature context for the benchmark IMF slopes used in this paper. Published high-mass IMF slopes from the compilation of \citet{hennebelle2024} are plotted against the IMF-depth proxy $1/f(m_\mathrm{lo})$, defined here as the number of total IMF stars expected per star above a study's lower-mass fitting limit $m_\mathrm{lo}$. Larger values therefore correspond to shallower surveys that constrain $\alpha$ using only the rare high-mass tail. Points are color-coded by broad environment class, and diamond markers denote measurements that include explicit binary corrections. Horizontal dotted lines mark the four benchmark $\alpha$ values used in this paper.}
\label{fig:imf_depth}
\end{figure}

Figure~\ref{fig:imf_depth} provides the literature context for these four benchmark $\alpha$ values taken from \citet{hennebelle2024}.\footnote{\url{https://github.com/mikegrudic/alphaplot}} We define the IMF-depth proxy $1/f(m_\mathrm{lo})$ as the number of total IMF stars expected per star above a study's lower-mass fitting limit $m_\mathrm{lo}$, so larger values correspond to shallower surveys that fit only the rare high-mass tail. Published high-mass slopes span a wide range across both fitted mass ranges and environments; the four $\alpha$ values used here are chosen to bracket that spread rather than to assert a unique environment-to-$\alpha$ mapping.

\begin{figure*}[t!]
\centering
\includegraphics[width=\textwidth]{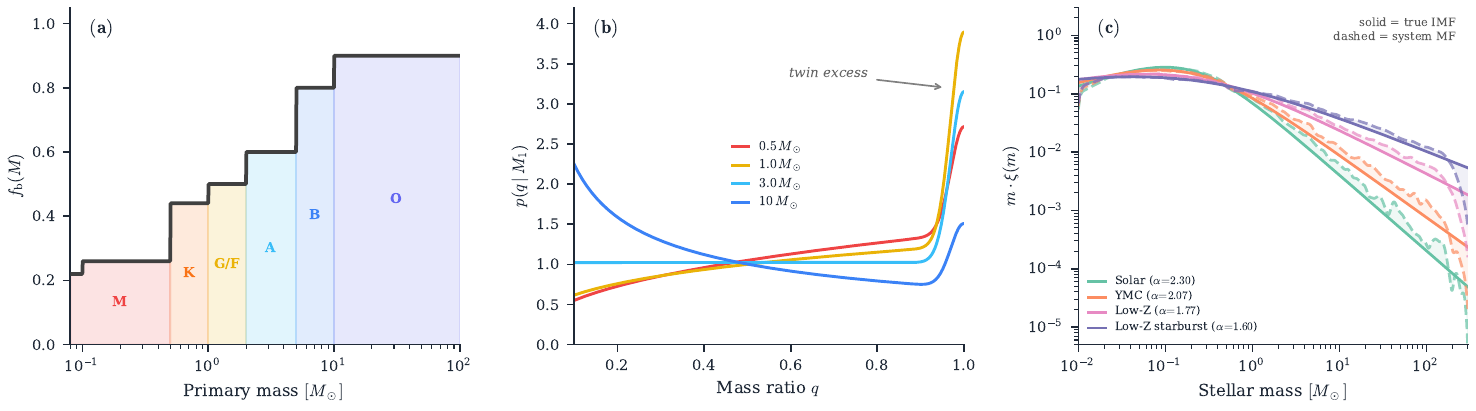}
\caption{Binary population model from \citet{moe2017}. \textbf{(a)} Mass-dependent binary fraction $\fb(\mpri)$, showing the step-function dependence from 22\% for late M~dwarfs to 90\% for O~stars. Colored bands mark spectral type regions (M, K, G/F, A, B, O). \textbf{(b)} Mass-ratio distribution $p(q \mid \mpri)$ for four representative primary masses spanning the full range of power-law index $\gamma$. The narrow peak at $q \approx 1$ is the ``twin excess.'' \textbf{(c)} Distortion of the observed system mass function under mass-addition (dashed) relative to the true single-star IMF (solid) for all four environments. The shaded region highlights the mass overestimate from unresolved binaries.}
\label{fig:binary_model}
\end{figure*}

\subsection{Binary Population Model} \label{sec:binaries}

We adopt a simplified, period-marginalized approximation of the \citet{moe2017} empirical binary population model. The full model characterizes multiplicity as a function of both primary mass and orbital period; we marginalize over period because our calculation depends on the integrated companion contribution, not the orbital parameters. This neglects the period-dependent mass-ratio distribution but is appropriate for our effective contamination model.

We also treat each system as either single or binary, ignoring higher-order multiples. This simplification is conservative for massive stars: because the true multiplicity frequency exceeds unity when triples and quadruples are common, a binary-only treatment underestimates the total unresolved contamination, likely biasing $\Ncross$ high. The misspecification test in \S\ref{sec:self-consistency} shows that assuming a flat mass-ratio distribution produces only small shifts in the recovered IMF slope in the two benchmark cases tested.

\vspace{6pt}\noindent\textbf{Effective binary probability.} For our single-or-binary forward model, we adopt an effective binary probability $\fb(\mpri)$ that approximates the incidence of unresolved companions with $q > 0.1$ as a function of primary mass, motivated by the multiplicity trends compiled by \citet{moe2017} and related reviews:
\begin{equation}
    \fb(\mpri) = \begin{cases}
    0.22 & \mpri < 0.1\,\msun \\
    0.26 & 0.1 \leq \mpri < 0.5\,\msun \\
    0.44 & 0.5 \leq \mpri < 1.0\,\msun \\
    0.50 & 1.0 \leq \mpri < 2.0\,\msun \\
    0.60 & 2.0 \leq \mpri < 5.0\,\msun \\
    0.80 & 5.0 \leq \mpri < 10.0\,\msun \\
    0.90 & \mpri \geq 10.0\,\msun
    \end{cases}
\end{equation}
This is not the multiplicity frequency of \citet{moe2017}, which can exceed unity for massive stars, but an effective binary probability calibrated to reproduce the strong mass dependence of stellar multiplicity \citep{duchene2013, offner2023, sana2025}. For massive stars, our binary-only treatment underestimates the contamination.

\vspace{6pt}\noindent\textbf{Mass-ratio distribution.} We approximate the \citet{moe2017} mass-ratio distribution after marginalizing over orbital period in our forward model as a power law plus a narrow twin excess:
\begin{equation}
    p(q \mid \mpri) \propto q^{\gamma(\mpri)} + \ftwin(\mpri)\,\mathcal{N}(q; 1, \sigma_\mathrm{twin}^2)
\end{equation}
where the effective slope $\gamma(\mpri)$ varies from $+0.4$ for M~dwarfs (favoring near-equal masses) to $-0.5$ for OB stars (favoring unequal masses). The twin excess is implemented as a truncated Gaussian near $q = 1$ with $\sigma_\mathrm{twin} = 0.03$ as a numerical convenience; the original \citet{moe2017} parameterization instead defines twins as an excess fraction for systems with $q > 0.95$, not a Gaussian. We adopt $\qmin = 0.1$ following \citet{moe2017}; this truncation is conservative, since the lowest-$q$ binaries contribute negligible bias under either operator ($\mobs \approx \mpri$ as $q \to 0$). Figure~\ref{fig:binary_model} illustrates these distributions.

\subsection{Observation Operators} \label{sec:obs}

The \emph{observation operator} $\mathcal{O}$ maps a stellar system to the scalar mass that would be inferred under a single-star interpretation:
\begin{equation}
    \mobs = \begin{cases}
    \mpri & \text{single star} \\
    \mathcal{O}(\mpri, m_2) & \text{unresolved binary}
    \end{cases}
\end{equation}
Conditioned on a system being unresolved, the inferred mass depends primarily on the component masses or luminosities, not directly on the orbital parameters. We compare two idealized observation operators to quantify the resulting bias.

\vspace{6pt}\noindent\textbf{Mass-addition operator.}
\begin{equation} \label{eq:mass-add}
    \mathcal{O}_M(\mpri, q) = \mpri(1 + q)
\end{equation}
The observer naively assigns the total system mass as the mass of a single star. An equal-mass binary ($q = 1$) is overestimated by 100\%. We treat this as an idealized benchmark rather than as a literal model of any single observing pipeline.

\vspace{6pt}\noindent\textbf{Luminosity-addition operator.}
\begin{equation} \label{eq:lum-add}
    \mathcal{O}_L(\mpri, q) = L^{-1}\!\left(L(\mpri) + L(q\mpri)\right)
\end{equation}
where $L(m)$ is the zero-age main sequence (ZAMS) mass--luminosity relation \citep{tout1996} and $L^{-1}$ is its inverse. This represents an idealized photometric case: the observer measures the combined luminosity perfectly and infers a mass assuming a single star. Because the ZAMS mass--luminosity relation is superlinear, the secondary contributes less to the inferred mass than to the total system mass. We compute $L^{-1}$ via Newton iteration on the \citet{tout1996} ZAMS luminosity, using the metallicity-dependent coefficients appropriate to each environment.

For unevolved ZAMS populations, mass-addition assigns a larger inferred mass than luminosity-addition, $m_1 + m_2 > L^{-1}(L_1 + L_2)$, for a superlinear mass--luminosity relation. In practice, multi-band color--magnitude diagram (CMD) fitting may reduce the bias further for unevolved populations, because color information can partially break the single-star/binary degeneracy \citep[e.g.,][]{liu2025pleiades}. However, evolved stars deviate from the ZAMS mass--luminosity relation, and evolved binaries can produce larger mass overestimates than our ZAMS calculation predicts (\S\ref{sec:hierarchy}). These results do not apply to evolved populations where the mass--luminosity relation deviates from the ZAMS; real surveys containing evolved stars require a dedicated treatment.

\subsection{Inference} \label{sec:inference}

\noindent\textbf{Naive model.} The single-star likelihood treats each observed mass $\mobs$ as an independent draw from the IMF:

\begin{equation}
    \ln \mathcal{L}_\mathrm{naive}(\alpha) = \sum_{i=1}^{N} \ln \xi(\mobs^{(i)} \mid \alpha)
\end{equation}

\noindent
This ignores the binary population entirely.

\vspace{6pt}\noindent\textbf{Binary-aware model.} Each observed mass may arise from either a single star or an unresolved binary. The mixture likelihood sums the single-star and binary contributions:

\begin{equation} \label{eq:binary-aware}
\begin{split}
    p(\mobs \mid \alpha) &= \bigl(1 - \fb(\mobs)\bigr)\,\xi(\mobs \mid \alpha) \\
    &\quad+ \!\int_{\mobs/2}^{\mobs/(1+\qmin)}\!
    \fb(\mpri)\,\xi(\mpri\!\mid\!\alpha) \\
    &\qquad \times g(q\!\mid\!\mpri)\,\frac{d\mpri}{\mpri}
\end{split}
\end{equation}

\noindent 
where $\xi(m \mid \alpha)$ is the IMF probability density normalized over the fitted mass interval. For single stars, $\mobs = \mpri$, so the singles contribution reduces to $(1 - \fb(\mobs))\,\xi(\mobs \mid \alpha)$. For binaries under the mass-addition operator, the observed mass satisfies $\mobs = \mpri(1+q)$. The binary contribution to the likelihood can therefore be written as

\[
\begin{aligned}
p_\mathrm{bin}(\mobs \mid \alpha)
&=
\int d\mpri \int dq\;
\fb(\mpri)\,\xi(\mpri \mid \alpha) \\
&\qquad \times g(q \mid \mpri)\,
\delta\!\bigl(\mobs - \mpri(1+q)\bigr).
\end{aligned}
\]

\noindent
At fixed $\mobs$ and $\mpri$, the implied mass ratio is

\begin{equation}
q^*(\mpri) = \frac{\mobs}{\mpri} - 1.
\end{equation}

\noindent
Using the delta-function identity $\delta(h(q)) = \delta(q-q^*)/|dh/dq|$ with $h(q)=\mobs-\mpri(1+q)$ gives $|dh/dq|=\mpri$, so the $q$-integral yields

\begin{equation}
p_\mathrm{bin}(\mobs \mid \alpha)
=
\int
\fb(\mpri)\,\xi(\mpri \mid \alpha)\,g(q^* \mid \mpri)\,
\frac{d\mpri}{\mpri}.
\end{equation}

\noindent
The integration limits follow from enforcing $q^*(\mpri) \in [\qmin, 1]$, giving $\mpri \in [\mobs/2,\; \mobs/(1+\qmin)]$. We evaluate the integral with 128-point Gauss--Legendre quadrature.

For the luminosity-addition operator, the same mixture logic applies, but both the inverse mapping and the Jacobian become more complicated because $\mobs$ is defined implicitly through the ZAMS mass--luminosity relation. The lower integration bound is $\mpri = L^{-1}(L_\mathrm{obs}/2)$, corresponding to the equal-mass case $q=1$. The upper bound is set by the minimum allowed mass ratio, $L(\mpri) + L(\qmin \mpri) = L_\mathrm{obs}$. In the implementation used here, we evaluate this upper bound with a numerical approximation,

\begin{equation}
\mpri^\mathrm{hi} \approx L^{-1}\!\left(L_\mathrm{obs} - L(\qmin\,\mobs)\right),
\end{equation}

\noindent
which differs from the exact solution by $< 0.07\%$ in relative primary mass across the full mass range ($1$--$100\,\msun$) for the steep ZAMS mass--luminosity relation and $\qmin = 0.1$ adopted here. The implied mass ratio is
\[
q^*(\mpri) = \frac{L^{-1}(L_\mathrm{obs} - L(\mpri))}{\mpri},
\]
and the Jacobian becomes $L'(\mobs)/({\mpri\,L'(q^*\mpri)})$, which follows from differentiating the implicit relation $L(\mobs) = L(\mpri) + L(q\mpri)$ with respect to $q$ and applying the inverse-function theorem for the monotonic ZAMS relation. This yields an approximate binary-aware luminosity-addition likelihood for the idealized ZAMS case. The binary population parameters ($\fb$, $\gamma$, $\ftwin$) are held fixed from \citet{moe2017} and are not inferred jointly with $\alpha$. The corresponding log-likelihood is

\begin{equation}
    \ln \mathcal{L}_\mathrm{bin}(\alpha) = \sum_{i=1}^{N} \ln p(\mobs^{(i)} \mid \alpha).
\end{equation}

Both models are inferred via Hamiltonian Monte Carlo \citep[HMC;][]{betancourt2018} using the No-U-Turn Sampler \citep[NUTS;][]{hoffman2014} as implemented in \numpyro{} \citep{phan2019composable, bingham2019pyro}. We run 2 chains of 1,000 samples each (after 500 warm-up samples) and verify convergence via $\hat{R} < 1.05$ and effective sample size $n_\mathrm{eff} > 100$. The prior on $\alpha$ is uniform on $[0.5, 4.0]$.

In the 640-run multi-seed campaign, 625 runs (98\%) achieve $\hat{R} < 1.01$ and $n_\mathrm{eff} > 460$ (median $n_\mathrm{eff} = 728$). The 13 severe convergence failures ($\hat{R} > 1.05$) are concentrated in the low-$\alpha$, large-$N$ corner, with lower sample quantiles near the prior boundary at $\alpha = 0.5$. All reported bias values and interquartile range (IQR) bands are computed from the convergence-filtered subset.

We define $\Ncross$ as the first tested sample size at which the absolute difference between the posterior median and the true slope exceeds the full equal-tailed 95\% credible-interval width, $Q_{0.975}-Q_{0.025}$. We use the individual scaling runs, without interpolation between tested sample sizes. This criterion identifies a regime in which the offset dominates the posterior uncertainty; it is more stringent than exclusion of the true slope from the 95\% credible interval.

All code is implemented in JAX \citep{jax2018github}, enabling automatic differentiation through the likelihood (required for HMC) and hardware acceleration on GPU. The binary population model is provided by \progenax{}, a JAX-native package for differentiable star cluster and stellar population initial conditions (Rosen, in preparation). The binary-aware likelihood evaluates a 128-point Gauss--Legendre quadrature integral per star, making it roughly two orders of magnitude slower per likelihood call than the naive model; this cost is acceptable for individual clusters and could be reduced for large-scale surveys via importance sampling, amortized inference, or lower quadrature order.

\section{Results} \label{sec:results}

\subsection{The Confidently Wrong Regime} \label{sec:confidently-wrong}

\begin{figure*}
\centering
\includegraphics[width=\textwidth]{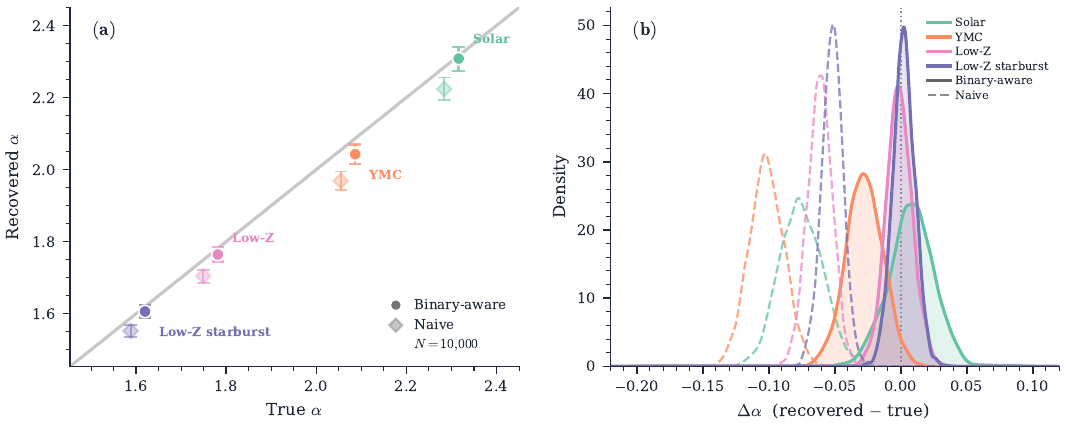}
\caption{Parameter recovery across four astrophysical environments at $N = 10{,}000$ using the mass-addition operator. \textbf{(a)} True vs.\ recovered $\alpha$: naive estimates (diamonds, with 95\% CIs) are systematically biased low, while binary-aware estimates (circles) are consistent with the truth (solid line). \textbf{(b)} Residual posterior distributions: naive posteriors (dashed) are shifted negative, confirming systematic bias; binary-aware posteriors (solid) are centered on zero.}
\label{fig:recovery}
\end{figure*}

We first test parameter recovery at $N = 10{,}000$ across all four environments using the mass-addition operator (Figure~\ref{fig:recovery}). The naive model systematically underestimates $\alpha$ in every environment, with single-realization biases of $\boldsymbol{-0.052}$ to $\boldsymbol{-0.102}$, while the binary-aware model substantially reduces the bias, with 95\% credible intervals containing the true $\alpha$ in three of the four environments. The true slope lies just above the upper interval limit for the YMC realization, consistent with the sampling variation expected across independent realizations.

\begin{figure*}
\centering
\includegraphics[width=\textwidth]{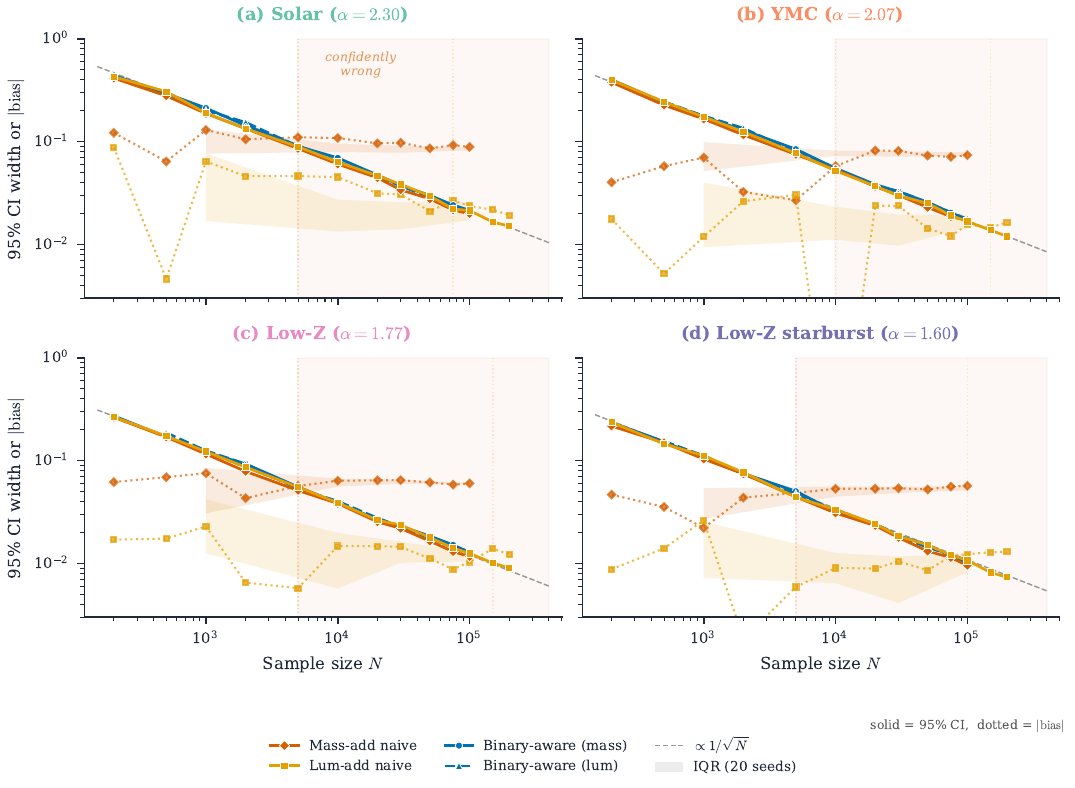}
\caption{The ``confidently wrong'' regime across four astrophysical environments. Each panel shows the 95\% credible interval width (solid lines) and absolute bias $|\alphanaive - \alphatrue|$ (dotted lines) as functions of sample size $N$ on log-log axes. Four inference configurations are compared: mass-addition naive (orange diamonds), luminosity-addition naive (amber squares), binary-aware mass-addition (blue circles, solid), and binary-aware luminosity-addition (blue triangles, dashed). The gray dashed line shows the $1/\sqrt{N}$ Bernstein--von Mises scaling. The shaded region begins at the first tested sample size where the absolute posterior-median offset for mass-addition exceeds the full 95\% credible-interval width. Both binary-aware models track the $1/\sqrt{N}$ scaling with bias consistent with zero.}
\label{fig:scaling}
\end{figure*}

\textbf{The statistical uncertainty shrinks with $N$, but the bias does not.} As sample size increases, the 95\% credible interval (CI) width shrinks as $1/\sqrt{N}$ for all models, but the absolute bias $|\alphanaive - \alphatrue|$ remains approximately constant (Figure~\ref{fig:scaling}). The crossover $\Ncross$ is the first tested sample size at which the absolute posterior-median offset exceeds the full 95\% CI width, as defined in Section~\ref{sec:inference}.

\begin{figure*}
\centering
\includegraphics[width=\textwidth]{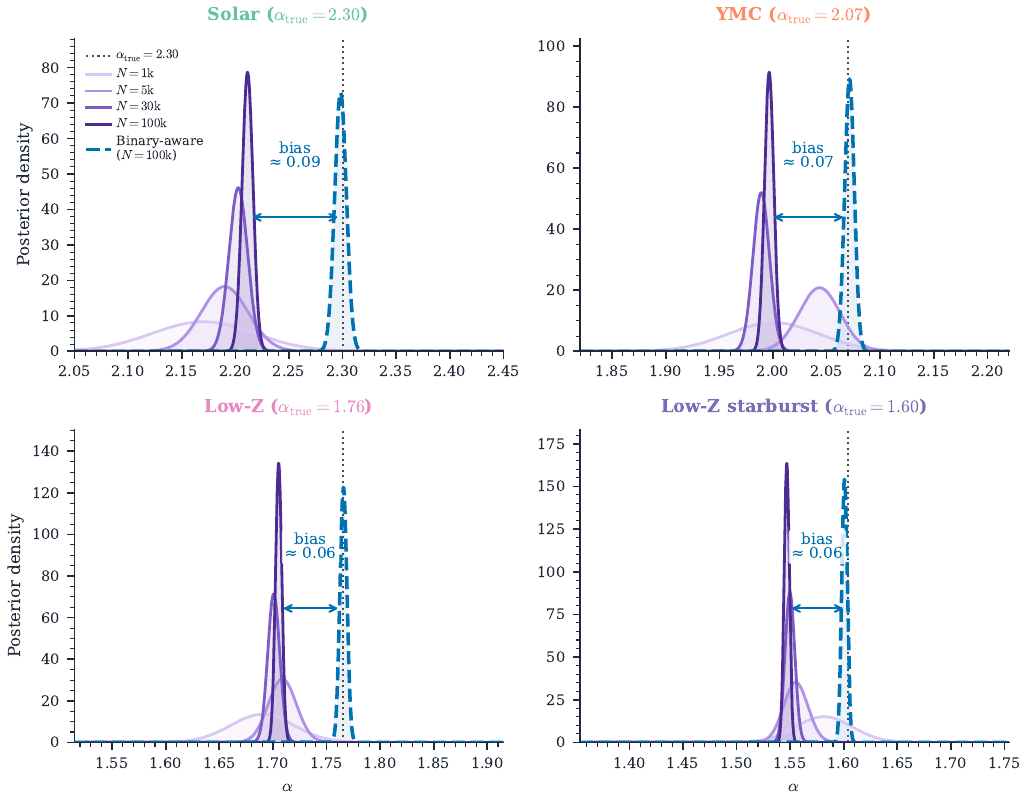}
\caption{The ``confidently wrong'' posteriors under mass-addition for all four environments. Naive posteriors (solid lines) at $N = 1{,}000$, $5{,}000$, $30{,}000$, and $100{,}000$ are shown with progressively darker shading; binary-aware posteriors (dashed) at $N = 100{,}000$ recover the truth (dotted vertical line) in every environment. As $N$ increases, naive posteriors narrow onto the biased value rather than the truth. Bias magnitudes decrease from Solar (0.086) to Low-$Z$ starburst (0.054), but all environments enter the confidently wrong regime. Posteriors are Gaussian approximations from HMC summary statistics.}
\label{fig:posteriors}
\end{figure*}

The naive posterior narrows onto the wrong value in every environment (Figure~\ref{fig:posteriors}). As $N$ increases from $1{,}000$ to $100{,}000$ under mass-addition, the posterior shrinks but converges away from the truth, while the binary-aware posterior at $N = 100{,}000$ recovers it. The 20-seed median bias is largest in the Solar environment (0.086); even in the Low-$Z$ starburst environment (0.054), the naive posterior at $N = 30{,}000$ already excludes the true $\alpha$.

Table~\ref{tab:scaling} summarizes the bias magnitudes and crossover sample sizes for both operators across all environments. These 20-seed median biases at $N = 100{,}000$ represent the converged systematic offset, as distinct from the single-realization recovery biases at $N = 10{,}000$.

\begin{deluxetable*}{lcccccccc}
\tablecaption{Scaling Analysis Summary \label{tab:scaling}}
\tablehead{
\colhead{Environment} & \colhead{$\alphatrue$} & \colhead{Mass-add} & \colhead{Mass-add} & \colhead{Mass-add} & \colhead{L-add} & \colhead{L-add} & \colhead{L-add} & \colhead{Delay} \\
\colhead{} & \colhead{} & \colhead{bias} & \colhead{CI ($N$=30k)} & \colhead{$\Ncross$} & \colhead{bias} & \colhead{CI ($N$=30k)} & \colhead{$\Ncross$} & \colhead{factor}
}
\startdata
Solar       & 2.300 & $-0.086$ & 0.034 & $\sim$5,000    & $-0.021$ & 0.038 & $\sim$75,000   & 15$\times$ \\
YMC         & 2.070 & $-0.076$ & 0.030 & $\sim$10,000   & $-0.017$ & 0.030 & $\sim$150,000  & 15$\times$ \\
Low-$Z$ GC  & 1.765 & $-0.060$ & 0.022 & $\sim$5,000    & $-0.011$ & 0.024 & $\sim$150,000  & 30$\times$ \\
Low-$Z$ starburst\tablenotemark{b}   & 1.604 & $-0.054$ & 0.018 & $\sim$5,000    & $-0.011$ & 0.018 & $\sim$100,000  & 20$\times$ \\
\enddata
\tablecomments{Bias values are 20-seed median asymptotic biases at $N = 100{,}000$. Naive CI widths are from single runs at $N = 30{,}000$. $\Ncross$ is the first tested sample size at which the absolute offset of the posterior median from the true slope exceeds the full 95\% credible-interval width. The delay factor is the ratio of L-add to mass-add crossover sample sizes.}
\tablenotetext{b}{Low-$Z$ starburst values use 14/20 (mass-add) and 17/20 (L-add) converged seeds at $N = 100{,}000$; the remaining seeds show $\hat{R} > 1.05$ and are excluded.}
\end{deluxetable*}

\textbf{The bias is approximately constant with $N$.} For both operators, the bias converges to a stable value by $N \gtrsim 2{,}000$ and remains constant thereafter. At small $N$ ($\lesssim 1{,}000$), the bias fluctuates and can occasionally flip sign due to sampling noise, but is not systematic.

\textbf{The bias magnitude depends on both operator and environment.} Mass-addition bias ranges from $-0.054$ (Low-$Z$ starburst) to $-0.086$ (Solar), while luminosity-addition bias ranges from $-0.011$ to $-0.021$. The luminosity-addition bias is 4--5$\times$ smaller than mass-addition across all environments. Steeper IMFs produce larger bias because the steeper decline at high masses makes the binary-shifted excess more prominent relative to the true single-star counts.

\textbf{The crossover sample size depends on the observation operator.} For mass-addition, all four environments enter the bias-dominated regime at $\Ncross \sim 5{,}000$--$10{,}000$. For luminosity-addition, the more optimistic operator, the crossover occurs at $\Ncross \sim 75{,}000$--$150{,}000$. This is 15--30$\times$ later than mass-addition.

\textbf{Binary-aware inference recovers the true slope.} The binary-aware model (blue curves in Figure~\ref{fig:scaling}) shows CI widths that shrink at the same $1/\sqrt{N}$ rate as the naive models, with residual bias fluctuating around zero. This shows that the \citet{moe2017} binary population model correctly accounts for the contamination in the matched-model case studied here.

The same holds under luminosity-addition. A binary-aware L-add likelihood using the same ZAMS mapping, Jacobian, and approximate upper integration bound described in Section~\ref{sec:inference} substantially reduces the bias in $\alpha$ across the tested sample sizes, $N = 500$--$100{,}000$ (blue dashed triangles in Figure~\ref{fig:scaling}). The residual bias is $|{\hat{\alpha} - \alpha_{\mathrm{true}}}| < 0.003$ at $N = 100{,}000$.

\textbf{The bias is robust across independent realizations.} To confirm that the bias is systematic rather than seed-dependent, we repeated the analysis with 20 independent random realizations per (environment, $N$, operator) combination --- 640 total runs, convergence-filtered (see Section~\ref{sec:inference}). The interquartile range (IQR) of the mass-addition bias across seeds is $\lesssim 0.007$ at $N = 100{,}000$, confirming that the bias is deterministic rather than a statistical fluctuation (Figure~\ref{fig:scaling}, shaded bands).

\textbf{Coverage fractions confirm the confidently wrong regime.} The coverage fraction, the fraction of 20 independent seeds whose 95\% CI contains the true $\alpha$, quantifies this directly (Table~\ref{tab:coverage}). For mass-addition, coverage drops to 0\% in all four environments at $N \geq 10{,}000$: every realization excludes the truth. For luminosity-addition, coverage degrades more slowly, reaching 0--5\% at $N = 100{,}000$.

\begin{deluxetable*}{lcccccc}
\tablecaption{Coverage Fractions by Operator, Environment, and Sample Size \label{tab:coverage}}
\tablehead{
\colhead{} & \multicolumn{3}{c}{Mass-addition} & \multicolumn{3}{c}{Luminosity-addition} \\
\cmidrule(lr){2-4} \cmidrule(lr){5-7}
\colhead{Environment} & \colhead{$N$=1k} & \colhead{$N$=10k} & \colhead{$N$=100k} & \colhead{$N$=1k} & \colhead{$N$=10k} & \colhead{$N$=100k}
}
\startdata
Solar       & 45\% & 0\% & 0\% & 85\%  & 80\% & 0\%  \\
YMC         & 60\% & 0\% & 0\% & 100\% & 80\% & 5\%  \\
Low-$Z$ GC  & 47\% & 0\% & 0\% & 90\%  & 74\% & 5\%  \\
Low-$Z$ starburst   & 63\% & 0\% & 0\% & 90\%  & 80\% & 0\%  \\
\enddata
\tablecomments{Fraction of 20 independent realizations whose 95\% credible interval contains the true $\alpha$, after convergence filtering ($\hat{R} < 1.05$). Here 0\% means that no converged realization covers the truth, while 5\% indicates a single covering realization after convergence filtering. A well-calibrated model would produce approximately 95\% coverage over repeated realizations. Mass-addition coverage drops to 0\% by $N = 10{,}000$ in every environment; luminosity-addition coverage persists longer due to the 4--5$\times$ smaller bias.}
\end{deluxetable*}

\subsection{Why Luminosity-Addition Bias is Smaller} \label{sec:lum-bias}

\begin{figure*}
\centering
\includegraphics[width=\textwidth]{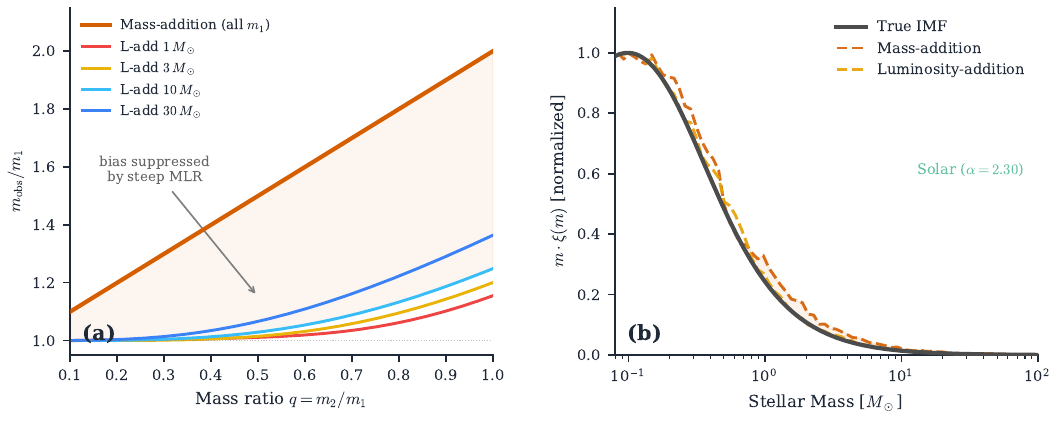}
\caption{Observation operator comparison. \textbf{(a)} Mass overestimate $\mobs/\mpri$ as a function of mass ratio $q$ for both operators. Mass-addition (orange) gives a universal linear relationship $1 + q$, while luminosity-addition (colored curves, one per primary mass) is strongly suppressed by the steep ZAMS mass--luminosity relation. The shaded region highlights the bias reduction from using photometric rather than mass-based methods. \textbf{(b)} Resulting system mass function distortion for the Solar environment ($\alpha = 2.30$): mass-addition (orange dashed) produces a larger departure from the true IMF (solid gray) than luminosity-addition (amber dashed), which sits close to the true distribution.}
\label{fig:operators}
\end{figure*}

The luminosity-addition bias is 4--5$\times$ smaller in magnitude than mass-addition because the steep ZAMS mass--luminosity relation $L \propto m^\eta$ (with $\eta \approx 3$--$4$ for $1$--$5\,\msun$, decreasing to $\eta \approx 2$--$3$ above $10\,\msun$) strongly suppresses the secondary's contribution to the inferred mass (Figure~\ref{fig:operators}). For a binary with mass ratio $q$:
\begin{equation}
    \frac{\mobs}{\mpri} = \left(1 + q^\eta\right)^{1/\eta} \approx 1 + \frac{q^\eta}{\eta} \quad \text{for } q \ll 1
\end{equation}
compared to $\mobs/\mpri = 1 + q$ for mass-addition. Even at $q = 0.5$ and $\eta \approx 3.5$ (typical for intermediate-mass ZAMS stars), luminosity-addition overestimates the mass by only $\sim 2.5\%$, versus 50\% for mass-addition. Since the \citet{moe2017} mass-ratio distribution peaks at low $q$ for high-mass stars ($\gamma < 0$), the majority of binaries contribute negligible bias under luminosity-addition.

The delay factor (ratio of L-add to mass-add $\Ncross$) is largest for the low-metallicity globular-cluster case (30$\times$) and smallest for the Solar-neighborhood and young-massive-cluster cases (15$\times$). Metallicity and IMF slope both vary across these environments, so we cannot separate their effects in this comparison.

\begin{figure*}
\centering
\includegraphics[width=\textwidth]{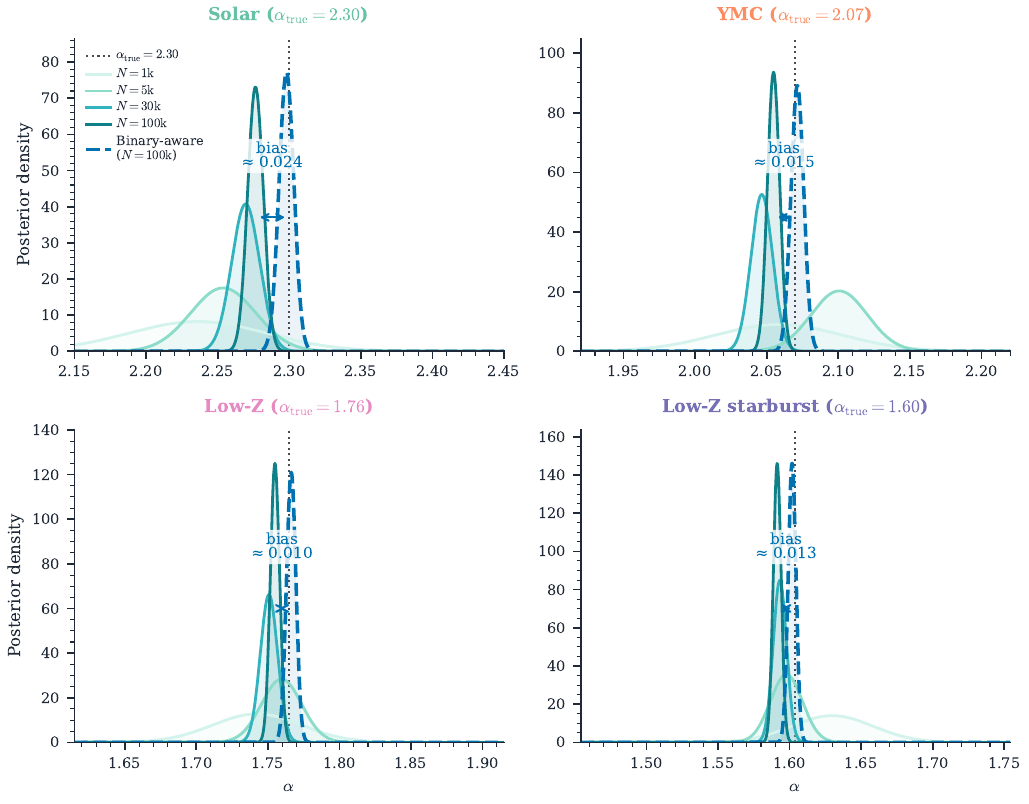}
\caption{Same as Figure~\ref{fig:posteriors}, but for the luminosity-addition operator. Naive posteriors (solid) use the \citet{tout1996} ZAMS to map system luminosity back to an apparent mass. The bias is 4--5$\times$ smaller than under mass-addition but follows the same pattern: narrowing posteriors that converge on the wrong value. Binary-aware posteriors (dashed) at $N = 100{,}000$ use the implemented L-add binary-aware likelihood and recover the truth in all four environments.}
\label{fig:posteriors-lum}
\end{figure*}

Figure~\ref{fig:posteriors-lum} illustrates this directly: under luminosity-addition, the naive posteriors at all four sample sizes cluster much closer to the true value than under mass-addition (compare Figure~\ref{fig:posteriors}), but the bias remains nonzero and the posteriors still converge on the wrong value at large $N$.

\section{Discussion} \label{sec:discussion}

Section~\ref{sec:results} shows that unresolved binaries produce a constant systematic bias on the inferred IMF slope that eventually dominates over statistical uncertainty. We now consider the implications for published measurements, the operator hierarchy for real surveys, and the limitations of our benchmark framework.

\subsection{Implications for Published IMF Slopes} \label{sec:implications}

A persistent bias of 0.01--0.09 has concrete consequences for published IMF slopes. For studies with $N \lesssim 5{,}000$, the bias is smaller than the statistical uncertainty and unlikely to dominate the error budget. For larger samples, it is not. The bias does not average away with increasing $N$ --- it is a systematic floor that every single-star analysis inherits.

The direction of the bias, naive analyses driving $\alpha$ too low, is consistent with the tendency of published high-mass slopes to scatter below the canonical Salpeter value $\alpha = 2.35$ \citep{hennebelle2024}. Environmental variations, measurement uncertainties, selection effects, and stellar-model choices all contribute to that scatter, so the literature offset cannot be attributed to binaries alone. But the binary contribution is the one systematic that grows in relative importance as statistical errors shrink. Re-analysis of archival data with binary-aware methods would help disentangle binary contamination from genuine environmental IMF variations.

Several recent analyses support the importance of binary treatment for IMF inference. Explicit binary corrections applied to Gaia DR2 yield a steeper slope than single-star analyses of the same data \citep{sollima2019}, a binary-aware analysis of the accreted Milky Way halo produces a bottom-heavy IMF \citep{hallakoun2021}, and \citet{wang2025solar} find that unresolved binaries must be accounted for when measuring the IMF in the solar neighborhood with Gaia DR3. In the SMC, \citet{legnardi2025} find using JWST photometry that binary contamination must be modeled to recover the intrinsic IMF \citep[see also][for photometric binary fractions in Magellanic Cloud clusters]{mohandasan2024}.

Other methods face related challenges. In 30~Doradus, VLT spectroscopy yields $\alpha \approx 1.90$ \citep{schneider2018aa}, one of the most compelling cases for a top-heavy IMF in a starburst-like environment. For LAMOST field stars, \citet{qiu2025field} independently simulate the effect of unresolved binaries on power-law indices and find that binary contamination must be corrected before interpreting metallicity-dependent IMF trends. Even dynamical evolution of paired subclusters can redistribute stellar masses and mimic IMF variations \citep{singhbal2025}. None of these studies constitutes a uniform test of the benchmark setup studied here, but collectively they indicate that binary treatment matters across diverse observational contexts.

\begin{figure}[t!]
\centering
\includegraphics[width=\columnwidth]{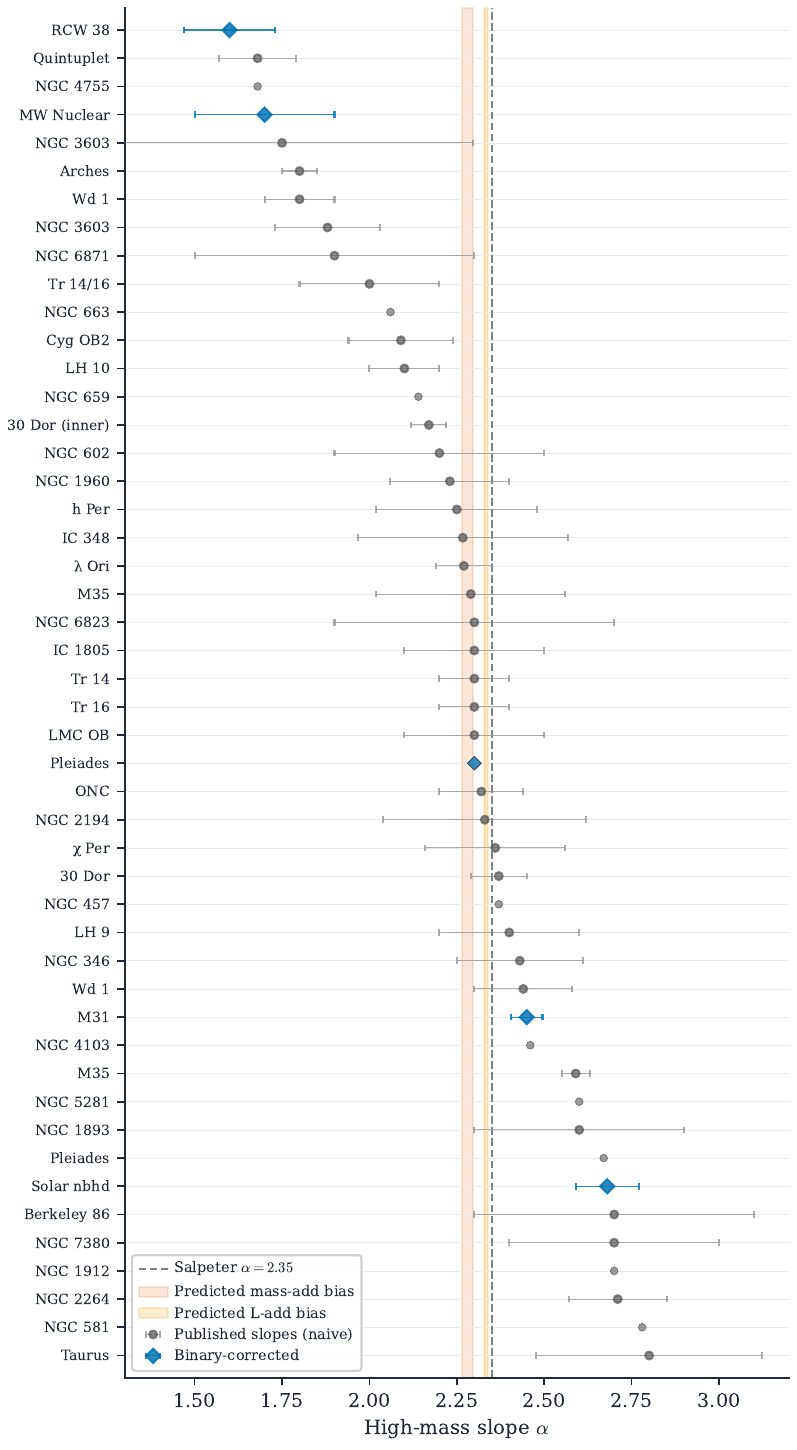}
\caption{Forest plot of 48 published high-mass IMF slopes from the \citet{hennebelle2024} compilation, sorted by slope value. Gray circles: slopes derived from single-star fitting (``naive''); blue diamonds: slopes corrected for binarity. Horizontal error bars show reported uncertainties where available. The vertical dashed line marks the Salpeter reference ($\alpha = 2.35$). Shaded bands show the predicted bias below Salpeter for mass-addition (orange) and luminosity-addition (amber) operators, spanning the range across our four environments. The comparison is illustrative; the compilation is too heterogeneous for causal interpretation.}
\label{fig:literature}
\end{figure}

Figure~\ref{fig:literature} places this prediction in context by compiling 48 published high-mass IMF slopes from the \citet{hennebelle2024} compilation. The compiled measurements span heterogeneous mass ranges, completeness limits, stellar models, and fitting methodologies, so this comparison is illustrative rather than a formal meta-analysis. The median published slope ($\alpha \approx 2.30$) is ${\sim}0.05$ shallower than Salpeter ($\alpha = 2.35$), an offset comparable in scale to the mass-addition bias found in our benchmark tests. Roughly half of the naive measurements have $\alpha$ values that fall within or below the predicted mass-addition bias range, but the compilation is too heterogeneous to support a stronger causal interpretation. Five measurements include explicit binary modeling \citep[following the classification of][]{hennebelle2024}; these scatter around the Salpeter value rather than systematically below it, though the sample is too small to draw strong conclusions.

Gaia DR3 and LSST provide opportunities to study binary contamination in large stellar samples \citep{gaia2023, ivezic2019}; early Rubin data already detect unresolved binary sequences in 47~Tucanae \citep{cordoni2025rubin}. Roman's wide-field near-infrared imaging will extend resolved-star censuses to dust-obscured young massive clusters where IMF slopes are most actively debated. The MiMO catalog \citep{li2025mimo} provides homogeneous mass function slopes for 1,232 open clusters from Gaia DR3 using a Bayesian framework that includes both single stars and unresolved binaries. \citet{malhotra2026} use simulation-based inference to jointly fit stellar masses and binary mass ratios for ${\sim}27{,}000$ stars in 42 Gaia open clusters, demonstrating the feasibility of binary-aware inference at scale.

\subsection{Operator Hierarchy and Real Surveys} \label{sec:hierarchy}

Our two operators provide idealized benchmarks for unresolved-binary bias in unevolved populations (\S\ref{sec:obs}). In practice, the relevant operator depends on the observation method:

\begin{itemize}
    \item \textbf{Luminosity-based mass estimates:} luminosity-addition maps the combined bolometric luminosity to a mass using an adopted mass--luminosity relation.
    \item \textbf{Multi-band CMD fitting} (isochrone matching): uses color information to help distinguish unresolved binaries from single stars \citep{liu2025pleiades}. Binaries with intermediate $q$ sit above the single-star sequence in the CMD, providing leverage to identify them \citep[e.g.,][]{milone2012, cordoni2023, liu2025pleiades}.
    \item \textbf{System masses treated as individual stellar masses:} mass-addition assigns the total binary mass to a single star.
    \item \textbf{Mass functions from luminosity functions} (applying a mass--luminosity relation to a luminosity histogram): approximately luminosity-addition.
\end{itemize}

\noindent Modern CMD-fitting pipelines that incorporate color information may reduce the bias relative to our luminosity-addition calculation for unevolved populations. Our crossover scales would then be conservative if the bias decreases relative to the posterior width.

Our luminosity-addition results use the ZAMS mass--luminosity relation \citep{tout1996} and therefore apply strictly to unevolved populations. Age spreads, differential extinction, and metallicity variations within a population are not included and can alter the masses inferred from photometry. Real clusters contain evolved stars where the mass--luminosity relation differs from the ZAMS, particularly near the main-sequence turnoff. A subgiant is more luminous than a main-sequence star of the same mass, so evolved binaries can produce larger mass overestimates than our ZAMS calculation predicts, potentially lowering $\Ncross$ if the slope bias increases relative to the posterior width. Clusters with significant evolved populations require a forward model incorporating post-main-sequence tracks.

Stellar rotation can broaden the main sequence and turnoff, and has been invoked to explain split main sequences above approximately $1.6\,\msun$ in young Magellanic Cloud clusters \citep{cordoni2018, milone2018}. Gravity darkening makes rotating stars' observed colors and brightnesses depend on viewing angle \citep{vonzeipel1924, espinosalara2011}. Rotation also changes stellar evolution and can extend main-sequence lifetimes, broadening the turnoff \citep{cordoni2018}. Globular clusters also contain populations with different helium and light-element abundances, which affect main-sequence colors \citep{milone2025}. \citet{milone2025} find that most binaries in their outer fields of 47 Tuc belong to the first population, whereas near the center, binary fractions differ only slightly between the first and second populations. Our binary model does not account for these abundance differences or variations in binary properties with population and radius.

\textbf{Practical guidance for observers.} When fitting IMF slopes to resolved stellar populations, compare the sample size with $\Ncross$ for a comparable observation model. In the idealized large-$N$ single-star-fitting regime studied here, mass-addition reaches the bias-dominated regime by $N \sim 5{,}000$--$10{,}000$, while ZAMS luminosity-addition reaches it by $N \sim 75{,}000$--$150{,}000$. In regimes where binaries are expected to matter and $N$ is large, a binary-aware likelihood is the principled solution. When that is not yet feasible, Table~\ref{tab:scaling} provides an order-of-magnitude benchmark for binary-driven systematics in our idealized setups.

This paper provides a benchmark framework for the large-$N$ single-star-fitting regime; survey-specific applications require forward models incorporating photometric noise, completeness, and stellar evolution. A full treatment for Gaia- and LSST-like photometry is in preparation.

\subsection{Model Self-Consistency} \label{sec:self-consistency}

Our analysis uses the \citet{moe2017} binary population model to both generate synthetic data and define the binary-aware likelihood. This tests recovery when the assumed binary model is correct, but does not establish robustness to model misspecification.

In reality, the true binary population may differ from the \citet{moe2017} parameterization. The \emph{naive bias is independent of the assumed binary model}: it depends only on the observation operator and the true binary population, and is present regardless of whether we correctly model it. The self-consistency concern applies only to the binary-aware recovery, not to the bias diagnosis.

To probe the sensitivity of the binary-aware recovery to model misspecification, we performed a limited misspecification test at $N = 10{,}000$ under mass-addition using three deliberately wrong binary models: (1) binary fractions scaled by 0.8 (underestimate by 20\%), (2) binary fractions scaled by 1.2 (overestimate by 20\%), and (3) a flat mass-ratio distribution replacing the mass-dependent \citet{moe2017} model. In the Solar environment, the shifts in recovered $\alpha$ relative to the correctly specified model are $-0.015$, $+0.013$, and $-0.007$, respectively, all $\lesssim 0.02$ and far smaller than the naive bias of $-0.108$. The same test in the Low-$Z$ starburst environment ($\alpha = 1.60$) yields shifts of $-0.011$, $+0.009$, and $+0.000$. In these two benchmark cases, even a 20\% error in the assumed binary fraction shifts the recovered $\alpha$ by only 10--15\% of the naive bias.

In these tests, the mixture likelihood responds smoothly to moderate perturbations in the assumed binary population, but the tests do not exhaust the misspecification space. For example, our period-marginalized approximation collapses the period-dependent mass-ratio structure documented by \citet{moe2017} into a single effective distribution; the sensitivity of the recovered $\alpha$ to this simplification has not been tested. The \citet{moe2017} model is calibrated primarily from Galactic field and cluster populations; environmental variations in binary properties, particularly in dense or low-metallicity systems, may shift the exact crossover scale but do not alter the qualitative result that unresolved binaries impose a systematic bias once statistical errors become small. We adopt field-star binary statistics as an empirical benchmark for quantifying how unresolved binaries bias IMF inference \citep{moe2017}. The globular-cluster measurements of \citet{milone2016} probe stars below the main-sequence turnoff and directly constrain binaries with $q>0.5$; comparing binary fractions therefore requires matched primary-mass and mass-ratio ranges. \citet{milone2016} find an approximately flat mass-ratio distribution for $q>0.5$, while our benchmark adopts a primary-mass-dependent distribution following \citet{moe2017}. Applying this framework to a specific cluster requires binary statistics appropriate to its stellar population. The model is empirically well-constrained against multiplicity surveys \citep{duchene2013}, so 20\% perturbations provide a reasonable first stress test.
\subsection{Limitations and Extensions} \label{sec:limitations}

\textbf{Higher-order multiples.} At least 15--20\% of OB stars are in triple or higher-order multiple systems \citep{duchene2013}. Our binary-only model therefore \emph{underestimates} the true bias: triples add additional mass contamination beyond what our binary model predicts. The binary-aware likelihood could be extended to include triples, but the binary contribution dominates the effect. Recent theoretical work linking the stellar IMF and multiplicity function as coupled outcomes of fragmentation \citep{thomasson2026} suggests that a self-consistent treatment of the IMF and binary population will be needed.

\textbf{Selection effects and completeness.} We assume a clean, complete sample with known mass bounds. In real surveys, binary boosting can scatter systems above a magnitude-limited detection threshold, inflating the apparent sample size. A full treatment of selection effects requires a survey-specific forward model, which we defer to a subsequent paper applying this framework to the LSST pipeline.

\textbf{Observational noise.} We assume perfect mass measurements (no photometric noise). Noise broadens the posterior (delaying the crossover to larger $N$) but also introduces additional systematic effects through nonlinear transformations of the error distribution. The net effect depends on the noise level and is survey-specific.

\textbf{Mass segregation.} In dynamically evolved clusters, mass segregation preferentially concentrates massive stars --- and their binary companions --- in the cluster core, where source crowding makes binary resolution harder. The effective binary contamination fraction for unresolved sources may therefore be spatially varying, with core-derived IMF slopes more biased than those from the cluster periphery.

\textbf{Multi-parameter inference.} We fix $\mu$ and $\beta$ to isolate binary contamination of the high-mass slope $\alpha$ from degeneracies among the IMF parameters.

In a joint inference of all three Maschberger parameters, the posterior could partially absorb the binary-induced distortion into the low-mass shape parameters, potentially reducing the apparent $\alpha$ bias but redistributing it across the parameter space. However, the bias is unlikely to vanish because the binary distortion is mass-dependent and asymmetric --- concentrated where the binary fraction is highest ($m \gtrsim 5\,\msun$) --- and this spectral shape, with ${\sim}7$ discrete break points in $\fb(\mpri)$, cannot be reproduced by any smooth two-parameter variation of $(\mu, \beta)$. A dedicated three-parameter analysis would clarify whether degeneracies between IMF shape and binary population parameters redistribute the bias.

\section{Conclusions} \label{sec:summary}

In this paper, we quantified the systematic bias that unresolved binaries impose on the inferred IMF high-mass slope $\alpha$ by comparing two idealized observation operators and testing recovery with a binary-aware mixture likelihood. We find that ignoring binaries produces a constant bias that statistical uncertainties eventually fail to cover --- a ``confidently wrong'' regime. Our main conclusions are:

\begin{enumerate}
    \item \textbf{Ignoring unresolved binaries is a systematic source of error in single-star IMF inference.} Across four astrophysical environments spanning $\alpha = 1.60$--$2.30$, the naive bias on $\alpha$ ranges from $-0.011$ to $-0.086$ depending on the observation operator.

    \item \textbf{Two observation operators quantify the bias in our benchmark tests.} Mass-addition ($\mobs = m_1 + m_2$) produces a bias magnitude of 0.054--0.086, compared with 0.011--0.021 for ZAMS luminosity-addition ($\mobs = L^{-1}(L_1 + L_2)$). The luminosity-addition bias is 4--5$\times$ smaller due to the steep mass--luminosity relation. Evolved populations can exceed the ZAMS luminosity-addition bias.

    \item \textbf{The crossover reaches survey-relevant sample sizes in these benchmark tests.} Mass-addition: $\Ncross \sim 5{,}000$--$10{,}000$. Luminosity-addition: $\Ncross \sim 75{,}000$--$150{,}000$. Both are well within LSST and Gaia capabilities.

    \item \textbf{Binary-aware inference recovers the true slope.} A mixture likelihood that marginalizes over the \citet{moe2017} binary population model substantially reduces the bias in $\alpha$ across the tested sample sizes (\S\ref{sec:self-consistency}).

    \item \textbf{Binary-aware inference is likely necessary for large-$N$ single-star analyses.} Current and upcoming datasets from Gaia, JWST, Roman, and Rubin/LSST will provide large stellar samples in rich clusters, where single-star IMF measurements can become bias-dominated unless binaries are modeled.

    \item \textbf{Limited misspecification tests remain favorable.} In mass-addition tests at $N = 10{,}000$ for two benchmark environments, perturbing the assumed binary fraction by $\pm 20\%$ or substituting a flat mass-ratio distribution shifts the recovered $\alpha$ by $\leq 0.02$, far smaller than the naive bias of 0.054--0.086.
\end{enumerate}

The IMF and binary population models used in this work are provided by \progenax{}, a differentiable JAX-native package for stellar population initial conditions (Rosen, in prep.). The next step is to extend this framework to synthetic color--magnitude diagrams that include photometric noise, completeness, and stellar evolution. We are developing \fluxax{}, a companion package for differentiable synthetic photometry, to enable this extension. Both packages will be described in future methods papers and released as open-source software.

\section*{Acknowledgments}
The author thanks Eric Sandquist and Kristen Dage for discussions about LSST-era star cluster science, and Mike Grudić for compiling the published IMF slope measurements used in Figures~\ref{fig:imf_depth} and~\ref{fig:literature}.
This work made use of \progenax{} (Rosen, in prep.),
JAX \citep{jax2018github},
NumPyro \citep{phan2019composable, bingham2019pyro},
matplotlib \citep{Hunter2007},
and NumPy \citep{harris2020array}.\\

\vspace{6pt}\noindent\textbf{AI Assistance Statement.} The author used Claude Code (Anthropic) and Codex (OpenAI) to assist with analysis scripting, numerical verification, and prose editing. All scientific content, methodology, and interpretations are the author's own. The author reviewed and verified all code, outputs, and text, and takes full responsibility for the integrity and accuracy of the work.

\vspace{6pt}\noindent\textbf{Data Availability.} The data underlying Figures 1--8 will be publicly available on Zenodo at publication (\dataset[doi:10.5281/zenodo.22983702]{https://doi.org/10.5281/zenodo.22983702}). The deposit includes documentation and plotting scripts. The analysis code and HMC posterior summaries are available from the author upon reasonable request. The \progenax{} package is planned for public release as open-source software.

\software{\progenax{} (Rosen, in prep.), JAX \citep{jax2018github}, NumPyro \citep{phan2019composable}, matplotlib \citep{Hunter2007}, NumPy \citep{harris2020array}}

\bibliographystyle{aasjournal}
\bibliography{references}

\end{document}